\documentclass[onecolumn]{cas-sc}

\usepackage[authoryear]{natbib}  
\usepackage{hyperref}  
\usepackage{graphicx}
\usepackage{natbib}
\usepackage{caption}
\usepackage{float}
\usepackage{subfigure}
\usepackage{xcolor}
\setcitestyle{numbers,square}

\def\tsc#1{\csdef{#1}{\textsc{\lowercase{#1}}\xspace}}
\tsc{WGM}
\tsc{QE}
\tsc{EP}
\tsc{PMS}
\tsc{BEC}
\tsc{DE}

\begin{document}
\let\WriteBookmarks\relax
\def\floatpagepagefraction{1}
\def\textpagefraction{.001}

\shorttitle{Nuclear Inst. and Methods in Physics Research, A}

\shortauthors{Mingzhe Yang et al.}

\title [mode = title]{Express Diagnostic of Intense Laser-driven MeV Radiation Source using Copper Isotopes}                      



%

\author[1]{Mingzhe Yang}


\fnmark[1]




\author[1]{Ziyao Wang}
\fnmark[1]

\author[1]{Jieru Ren\corref{cor1}}
\fnmark[1]
\cormark[1]
\ead{renjieru@xjtu.edu.cn}

\author[1]{Wenqing Wei}

\author[1]{Benzheng Chen}

\author[1]{Bubo Ma}

\author[1]{Shizheng Zhang}

\author[1]{Lirong Liu}

\author[1]{Fangfang Li}

\author[1]{Jie Xiong}

\author[1]{Hongwei Yue}

\author[1]{Zeyu Lai}

\author[1]{Wenxuan Li}

\author[1]{Dieter H.H. Hoffmann}

\author[2]{Olga N. Rosmej}

\author[3]{Parysatis Tawana}

\author[4]{N. E. Andreev}

\author[4]{I. R. Umarov}

\author[5]{Zhigang Deng}

\author[5]{Wei Qi}

\author[5]{Shaoyi Wang}

\author[5]{Quanping Fan}

\author[5]{Zongqiang Yuan}

\author[5]{Weiwu Wang}

\author[5]{Bo Cui}

\author[5]{Xiaohui Zhang}

\author[5]{Yuchi Wu}

\author[5]{Weimin Zhou}

\author[5]{Jingqin Su}

\author[6]{Rui Cheng}

\author[1]{Yongtao Zhao}
\fnmark[1]
\cormark[1]
\ead{zhaoyongtao@xjtu.edu.cn}

\affiliation[1]{organization={MOE Key Laboratory for Nonequilibrium Synthesis and Modulation of Condensed Matter, School of Physics, Xi'an Jiaotong University},
	addressline={},
	city={Xi'an},
	postcode={710049},
	country={China}}

\affiliation[2]{organization={GSI Helmholtzzentrum für Schwerionenforschung},
	addressline={},
	city={Darmstadt},
	postcode={64291},
	country={Germany}}

\affiliation[3]{organization={Institute of Optics and Quantum Electronics (IOQ), University Jena},
	addressline={},
	city={Jena},
	postcode={07743},
	country={Germany}}
	
\affiliation[4]{organization={Joint Institute for High Temperatures, Russian Academy of Sciences},
	addressline={},
	city={Moscow},
	postcode={125412},
	country={Russia}}

\affiliation[5]{organization={
		National Key Laboratory of Plasma Physics, Laser Fusion Research Center, China Academy of Engineering Physics},
	addressline={},
	city={Mianyang},
	postcode={621900},
	country={China}}

\affiliation[6]{organization={Institute of Modern Physics, Chinese Academy of Sciences},
	addressline={},
	city={Lanzhou},
	postcode={730000},
	country={China}} 

\cortext[cor1]{Corresponding author}



\begin{abstract}
We explored the generation and diagnosis of high-brightness MeV bremsstrahlung radiation caused by intense beam of relativistic electrons propagating in a tantalum converter. The intense electron beam was produced through direct laser acceleration mechanism in the interaction of relativistic high-power sub-ps laser pulse with near critical density plasma. We propose to detect the divergence angle and photon fluence of high-brightness and high-energy $\gamma$ radiation source based on the nuclear activation method. The radioactive \(^{62}_{29}\text{Cu}\) was generated through photonuclear reactions \(^{63}_{29}\text{Cu}\)($\gamma$,n) \(^{62}_{29}\text{Cu}\) and the subsequent $\beta^+$ decay of \(^{62}_{29}\text{Cu}\) was measured to derive characteristics of the $\gamma$ radiation source.
This method provides an express approach to diagnose the laser-driven MeV radiation source and a potential efficient way to produce \(^{62}_{29}\text{Cu}\) isotopes.
\end{abstract}


\begin{highlights}
\item Novel Rapid Method for Detecting MeV Gamma Radiation Sources
\item Efficient Production of High-Activity Isotopes
\item Mechanism for Generating a Stable, High-Brightness Radiation Source
\end{highlights}


\maketitle
\footnote{Ziyao Wang contributed equally to this work.}

\section{Introduction}

Laser-driven relativistic electron beams are characterized by short pulse duration, high energy and high current. They find applications across various scientific domains, particularly in the generation of bright, high-energy $\gamma$ radiation sources \cite{andreev2023}. $\gamma$ beams play a crucial role in inducing nuclear reactions, serving as an essential tool to investigate the properties of excited states in atomic nuclei \cite{veres1973}. Furthermore, they offer a pathway to explore the fundamental principles underlying nuclear excitation modes \cite{giulietti2008}. This groundbreaking research on photon-nuclear reactions was first performed by James Chadwick and Maurice Goldhaber, who used photons with energies greater than 2.2 $MeV$ to disintegrate deuterium \cite{chadwick1934}.
There is a high abundance of $\gamma$ rays in the universe, and these high-energy photons contribute to the synthesis of elements \cite{hoffmann1983}.  In addition, photon-nuclear reactions can also be employed for isotope production \cite{ruth2009}. For example, Ari Foley successfully produced nuclear forensic isotopes using the interaction of bremsstrahlung X-rays with \textsuperscript{238}U and \textsuperscript{232}Th \cite{foley2021}.

Limited by the low intensity and energy of $\gamma$ radiation source in the last century, and the majority of photonuclear reaction cross-sections (\(\gamma, n\)) are on the order of millibarns only \cite{koning2019}.
Isotope yields from photonuclear reactions driven by high-energy $\gamma$ radiation are generally inferior to those produced by conventional accelerators with the \((p, n)\) and \((\alpha, n)\) reaction channels \cite{desaavedra2018}.
For this reason, global interest in photonuclear methods has gradually waned since the 1970s. 
Hereafter, researchers tried to enhance the photonuclear reaction yields via modifying the targets like using stacks \cite{aizatsky2010}, proposing to use kinematic recoil methods\cite{starovoitova2014}, and increasing both the intensity and energy of the electron source and consequently the $\gamma$ sources \cite{aliev2019}. 
Typically in the experiment performed in 2010 \cite{aizatsky2010}, a production of  \(^{67}_{29}\text{Cu}\) based on photonuclear method reached activity as high as 5.6 $GBq$, and a possible yield up to 0.89 $MBq$/($\mu$$A*h$) \cite{kazakov2021}. 

With the development of high-power laser technology, photonuclear reaction has once again become a cutting-edge research area.  
Based on high-power laser technology, electron beams can be efficiently accelerated to energies ranging from several MeV up to some of GeV through mechanisms including direct laser acceleration (DLA), laser wakefield acceleration (LWFA) and so on.
The electron beam generated via the DLA mechanism carries a charge approaching the microcoulomb level, reaches energies up to 100 $MeV$, and exhibits good collimation \cite{pukhov1999,pukhov2002,tavana2023}.
High-energy and high-current electron beams can be applied to produce high-brightness  $\gamma$ source. Such sources of $\gamma$ radiation play an indispensable role in medical imaging \cite{nakel1994}, industrial detection \cite{suliman2016}, astrophysics \cite{chluba2020} and nuclear research \cite{metag2012}. Taking advantage of of high-power lasers, it has become feasible to generate high-activity isotopes through photonuclear reactions \cite{ma2019}. 

In recent years, research on foam targets attached a lot of attention due to its high conversion efficiency of energy from laser to particles.
It has been demonstrated that foam target can improve the intensity of high current electron beams \cite{jiang2023,rosmej2020}.
Specially, our co-author Rosmej used a picosecond laser \cite{bagnoud2010} with intensity of \(10^{19} \, \mathrm{W/cm^2}\) to hit sub-mm thick foam layer converted into  plasma of near critical electron density (NCD) by an additional ns-pulse. The critical plasma density for laser light propagation is the density at which the plasma frequency equals the laser's frequency. When the plasma density reaches this critical value, the laser light is reflected rather than propagated through the plasma. 
For a laser with wavelength of \( 1~\mu\mathrm{m} \), the corresponding critical electron density is approximately \(10^{21} \, \mathrm{/cm^3}\). 
This resulted in the generation of an electron beam with a charge on the order of micro-coulombs and energies above hundred MeV. Upon transmitting this beam through a 1 $mm$-thick gold converter, a high-brightness secondary $\gamma$ radiation source and a neutron-production source were realized. \cite{gunther2022}. 
At PHELIX experiment, High-Purity Germanium (HPGe) detector was used to obtain the energy spectrum of secondary $\gamma$ radiation source.
However, this approach was time-consuming and the use of HPGe detector required a low‑background environment. In addition, it cannot measure the divergence angle of the $\gamma$ radiation source.

In this work, we generated the NCD plasma by indirectly heating the foam target with soft X-rays generated in a gold hohlraum using a 150 J ns pulse \cite{su2015}. After a few nanoseconds, the foam transformed into plasma, which was hydrodynamically more stable than when the foam was directly irradiated with a ns pulse \cite{rosmej2020,rosmej2019}.
More importantly, we presented a convenient approach to detect the divergence angle and photon fluence of high-brightness and high-energy $\gamma$ radiation source at energy around 17 $MeV$ based on the copper activated method.
In our experiment, interaction of relativistic laser pulse with long-scale NCD plasma resulted into production of directed electron beams with energy and charge 10 times higher than by irradiation of conventional foils. 
Propagation of such electrons within  materials with high atomic number leads to generation of directed bremsstrahlung with photon energies in the range of giant dipole resonance (GDR) and above. 
The energy of the $\gamma$ radiation source obtained in our experiment surpasses the activation sheet's photonuclear reaction threshold, driving photonuclear reactions. We employed copper activation as a diagnostic tool to quickly determine the divergence angle and photon fluence of the $\gamma$ radiation source in the energy of 17.5 $\pm$ 2.5 MeV.

\section{Experimental Setup}

Our experiments were conducted at the XG-III laser facility of Laser Fusion Research Center in Mianyang. The experimental setup is depicted in Fig.1. 
Here we used a nanosecond laser with an energy output of 120 $J$ to irradiate the gold hohlraum and produced soft X-rays. The soft X-rays heat the foam target (C\textsubscript{9}H\textsubscript{16}O\textsubscript{8}, density of 2 mg/cc and thickness of 1 mm) to generate NCD plasma state \cite{rosmej2015}. 
The approach of producing NCD plasma by heating foams with hohlraum radiation was widely used and had good repetition \cite{ren2020,ren2023,ma2021}.

When the NCD plasma was fully formed after 9 nanoseconds, a picosecond laser interacted with the NCD plasma, generating high-current, high-energy and well directed electrons. This picosecond laser pulses has a duration of 800 femtoseconds, a focus spot smaller than \( 20~\mu\mathrm{m} \) and a maximum intensity of \(  \ 10^{19} \) W/cm$^2$. It can deliver energies up to 150 $J$ and more than 49\% of the laser energy can be converted. 
The interaction of these well-directed, high-current DLA electrons with a 3 mm thick tantalum converter produced the MeV $\gamma$ radiation source through bremsstrahlung mechanism.
A copper sheet with the size of 50 $mm$ * 50 $mm$ * 0.05 $mm$  was placed 5 $cm$ away from the target along the laser axis, and it was used to detect the MeV bremsstrahlung based on the nuclear activation method.
Several other detectors were positioned behind the copper sheet, so the central slit was cut in the copper sheet.
\begin{figure}[!htbp]
    \centering
    \subfigure[]{
        \includegraphics[scale=.5]{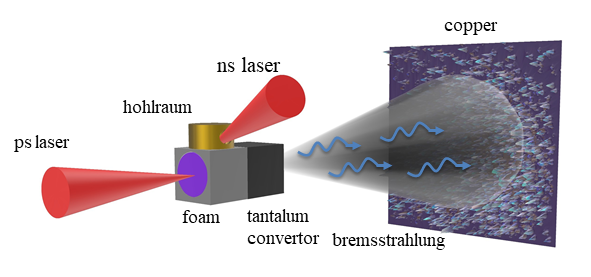}
        \label{fig:subfig1}
    }
    \subfigure[]{
        \includegraphics[scale=.5]{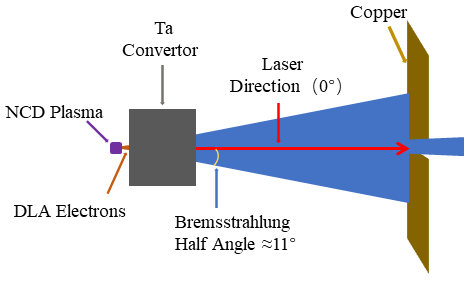}
        \label{fig:subfig2}
    }
    \caption{(a) Experimental setup diagram: A nanosecond laser irradiates the gold hohlraum converter to generate to generate soft X-rays, which in turn heat TCA foam and convert it into NCD plasma, producing the NCD plasma.  A picosecond laser is focused onto the NCD plasma target, generating DLA electrons. The Ta plate of 3 mm thickness is tightly attached to the back of the NCD plasma target. The copper is 5 centimeters away from the target; (b) Side view of the experiment layout illustrating the bremsstrahlung generation and half divergence angle.}
    \label{fig:main}
\end{figure}
When the $\gamma$ radiation in the GDR energy region irradiated the copper sheet,  \({}^{64}_{29}\text{Cu}\), \({}^{62}_{29}\text{Cu}\) and \({}^{61}_{29}\text{Cu}\) were produced. These isotopes are unstable and undergo decay, emitting positrons, electrons, and photons.
The Image Plate (Fuji-SR) was used to detect the particles emitted from the activated copper sheet \cite{yang2017}.
We used the activated area of the copper sheet to estimate the divergence angle, and total number of emitted particles to determine the fluence in the energy range of 17.5 $\pm$ 2.5 $MeV$ of the $\gamma$ radiation source.
During the process of measuring the particles emitted from the radioactive isotopes, the IP was tightly attached to the irradiated copper sheet in a dark room.
The IP was irradiated for 15 minutes, covering roughly two half-lifes of \(_{29}^{62}\mathrm{Cu}\). 
After every shot, we used a new copper sheet. 

The copper sheet was natural copper composed of 69\%  \({}^{63}_{29}\text{Cu}\) and 31\% \({}^{65}_{29}\text{Cu}\), and all of them are stable. Using TALYS-1.97 \cite{tayls2025} program, we obtained the photonuclear reaction cross-sections of \({}^{63}_{29}\text{Cu}\) (\(\gamma, n\)) \({}^{62}_{29}\text{Cu}\), \({}^{65}_{29}\text{Cu}\) (\(\gamma, n\)) \({}^{64}_{29}\text{Cu}\) and \({}^{63}_{29}\text{Cu}\) (\(\gamma, 2n\)) \({}^{61}_{29}\text{Cu}\), \({}^{65}_{29}\text{Cu}\) (\(\gamma, 3n\)) \({}^{62}_{29}\text{Cu}\), as illustrated in Fig. 2.
The figure clearly shows that the energy threshold for single neutron emission \({}^{63}_{29}\text{Cu}\) (\(\gamma, n\)) \({}^{62}_{29}\text{Cu}\), \({}^{65}_{29}\text{Cu}\) (\(\gamma, n\)) \({}^{64}_{29}\text{Cu}\) are lower than that for multi-neutron emission \({}^{63}_{29}\text{Cu}\) (\(\gamma, 2n\)) \({}^{61}_{29}\text{Cu}\), \({}^{65}_{29}\text{Cu}\) (\(\gamma, 3n\)) \({}^{62}_{29}\text{Cu}\). 
The cross section of \({}^{63}_{29}\text{Cu}\) (\(\gamma, n\)) \({}^{62}_{29}\text{Cu}\), \({}^{65}_{29}\text{Cu}\) (\(\gamma, n\)) \({}^{64}_{29}\text{Cu}\) are also much higher than \({}^{63}_{29}\text{Cu}\) (\(\gamma, 2n\)) \({}^{61}_{29}\text{Cu}\), \({}^{65}_{29}\text{Cu}\) (\(\gamma, 3n\)) \({}^{62}_{29}\text{Cu}\).
The bremsstrahlung energy spectrum exponentially decreases with increasing energy of photons \cite{seltzer1985}.
Therefore, the multiple neutron emission reaction channels of \({}^{63}_{29}\text{Cu}\) (\(\gamma, 2n\)) \({}^{61}_{29}\text{Cu}\), \({}^{65}_{29}\text{Cu}\) (\(\gamma, 3n\)) \({}^{62}_{29}\text{Cu}\) are neglected for the later analysis of copper activation yield and distribution.
\begin{figure}[!htbp]
	\centering
	\includegraphics[scale=.4]{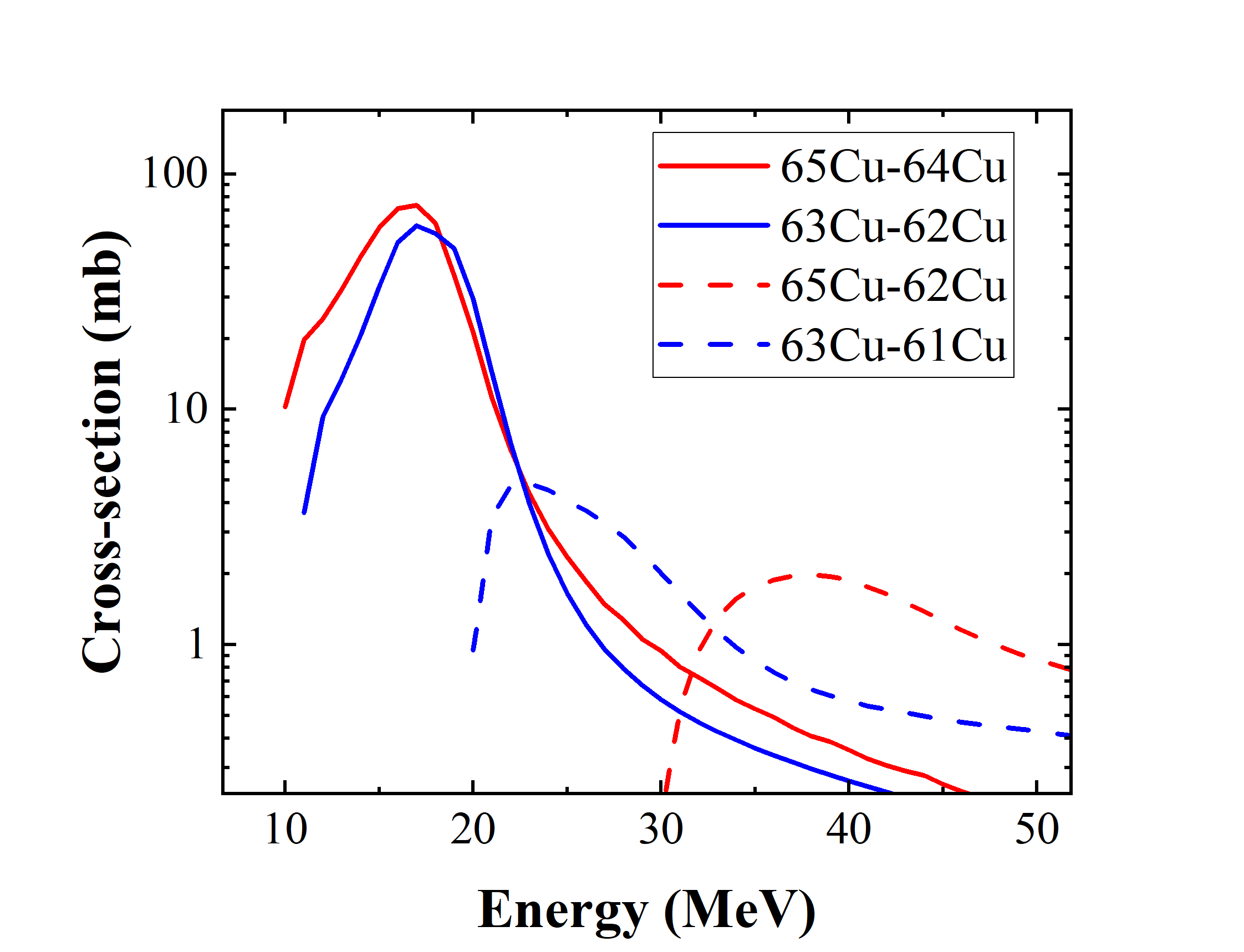}
	\caption{The cross sections of \({}^{63}_{29}\text{Cu}\) (\(\gamma, n\)) \({}^{62}_{29}\text{Cu}\), \({}^{65}_{29}\text{Cu}\) (\(\gamma, n\)) \({}^{64}_{29}\text{Cu}\) and \({}^{63}_{29}\text{Cu}\) (\(\gamma, 2n\)) \({}^{61}_{29}\text{Cu}\), \({}^{65}_{29}\text{Cu}\) (\(\gamma, 3n\)) \({}^{62}_{29}\text{Cu}\), which were calculated using TALYS-1.97.}
	\label{FIG:2}
\end{figure}

The radioactivate isotopes \({}^{62}_{29}\text{Cu}\) and \({}^{64}_{29}\text{Cu}\) emitted positrons, electrons and photons.
The IP-measured signal originated mainly from the positrons decayed from \(_{29}^{62}\mathrm{Cu}\). 
The reasons were as follow.
We neglected the response of IP to photons, whose maximum energy was 511 $keV$ mainly came from the positron annihilation, for two reasons.
First, the IP has more than 100 times higher sensitivity to positrons compared to 511 keV photons \cite{bonnet2013}. 
Second, the positron energy emitted from \(_{29}^{62}\mathrm{Cu}\) is 1.3 $MeV$ \cite{nndc2021} and its stopping path in copper \cite{shiroka2006}, after which annihilation can occur, is significantly larger than the copper sheet thickness we used.
In this case, positron scattering can be neglected as well.
\begin{table}[H]
	\caption{Decay information of \({}^{64}_{29}\text{Cu}\) and \({}^{62}_{29}\text{Cu}\) from NIST database.}
	\label{tab:fixed}
	\begin{tabular}{|c|c|c|c|c|c|c|}
		\hline
		\textbf{Parent Nucleus} & \textbf{Half Life \(T_{1/2}\)} & \textbf{Decay Mode} & \textbf{Positron Energy (keV)} & \textbf{Intensity}  \\
		\hline
        \(_{29}^{64}\mathrm{Cu}\) & 12.7006 hours & $\epsilon$ + $\beta^{+}$ = 61.5\% & 278.009 & 17.49\%  \\
		\(_{29}^{62}\mathrm{Cu}\) & 9.67 minutes & $\epsilon$ + $\beta^{+}$ = 100\% & 1320.7  & 97.6\%  \\
		\hline
	\end{tabular}
\end{table}
We also neglected the contribution of \({}^{64}_{29}\text{Cu}\). Because of the similar cross section of \({}^{63}_{29}\text{Cu}\) (\(\gamma, n\)) \({}^{62}_{29}\text{Cu}\) and \({}^{65}_{29}\text{Cu}\) (\(\gamma, n\)) \({}^{64}_{29}\text{Cu}\), we estimated the yield ratio of \({}^{62}_{29}\text{Cu}\) and \({}^{64}_{29}\text{Cu}\) was to be 7:3, corresponding to the abundance of \({}^{63}_{29}\text{Cu}\) and \({}^{65}_{29}\text{Cu}\). 
Referring to NIST database \cite{nist2025}, we got the half-lives, decay modes, characteristic positron energies, and the associated emission probabilities of \({}^{64}_{29}\text{Cu}\) and \({}^{62}_{29}\text{Cu}\), which are summarized in Table 1.
The half-lives are 9.67 minutes for \({}^{62}_{29}\text{Cu}\) and 12.7 hours for \({}^{64}_{29}\text{Cu}\), with positron-emission probabilities of 97.6$\%$ and 17.49$\%$, respectively. What's more, the electron decay branching ratio is about twice that of the positron decay from \({}^{64}_{29}\text{Cu}\).
We measured the activated copper sheet in the time slot of 12 minutes to 27 minutes after the shots. Over this 15 minutes, the number of the positrons and electrons from the decay of \({}^{64}_{29}\text{Cu}\) amounts to less than 1$\%$ of \({}^{62}_{29}\text{Cu}\).

\section{Results and discussion}
\subsection{Divergence angle}
We used the activated area of copper sheet to determine the divergence angle of the $\gamma$ radiation source. The \(_{29}^{62}\mathrm{Cu}\) undergone $\beta^+$ decay and emitted positrons. These positrons were recorded by the IP \cite{doyama1997}. We used the IP reader to acquire the gray-scale value $G$ per pixel caused by positrons \cite{bonnet2013,tanaka2005}. The spatial distributions of gray-scale value intensity over the full area of the copper sheets from shot 102 and shot 105 were presented in Fig. 3.

\begin{figure}[!htbp]
	\centering
	\includegraphics[scale=.4]{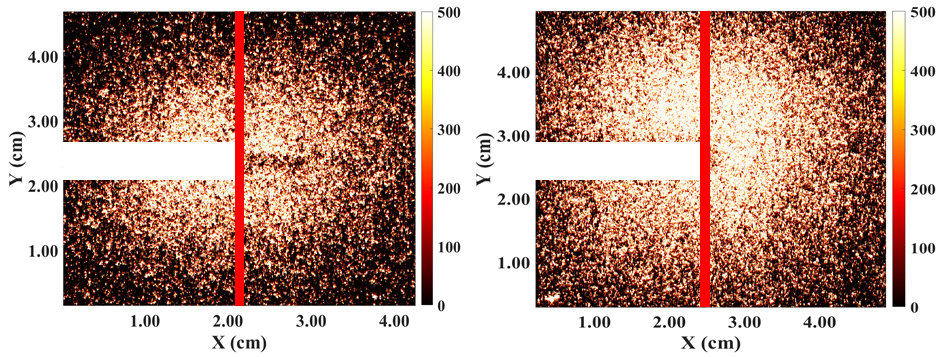}
	\caption{The spatial distributions of gray-scale value intensity over the full area of the copper sheets from shot 102 and shot 105. The white areas represent regions where the copper sheet were cut, indicating no signal. The red rectangular region indicated the area among which we made the statistics for the angular distribution analysis.}
	\label{FIG:3}
\end{figure}

We made the statistics of selected red area as shown in Fig. 3 to derive the divergence angle of the $\gamma$ radiation source.
The blank areas were filled using the gray-scale values of the surrounding regions. 
The spatial scale distribution was converted into the angle scale of the $\gamma$ radiation source based on the distance between the copper sheet and the target.
The number of positrons $N$ was determined using Eq. (1) and Eq. (2), as explained in section 3.2.
The angle distribution of the positrons on the IP is illustrated in Fig. 4. The full width at half maximum (FWHM) of the curve was 22.4$^\circ$ for shot 105 and 20.7$^\circ$ for shot 102, representing the full divergence angle of the $\gamma$ radiation source. In shot 102, laser was focused on the surface of the NCD plasma, whereas in shot 105, it was focused 200 \(\mu\)m inside of the NCD plasma. 
The uncertainty is about 4$^\circ$ from the emission angle of positrons.
\begin{figure}[!htbp]
	\centering
	\includegraphics[scale=.3]{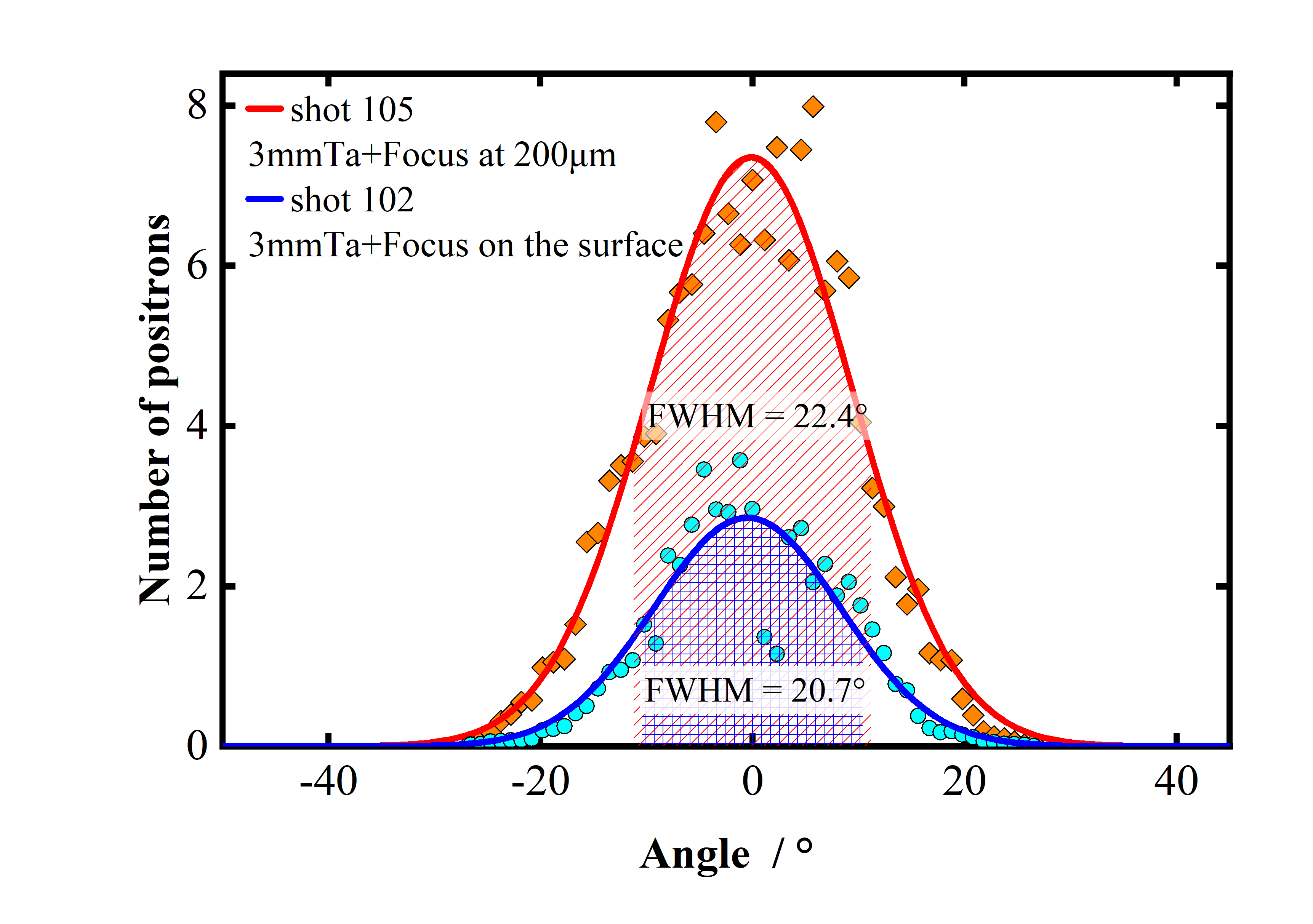}
	\caption{Results of the angular distribution of positrons on the IP for different laser focus positions in shot 102 and 105.}
	\label{FIG:4}
\end{figure}
This average half divergence angle of about 11$^\circ$ is much smaller than the result from NOVA \cite{stoyer2001}, where their divergence angle was about 105$^\circ$. We attribute this to the contribution of collimated relativistic electrons beam characteristic caused by the DLA process in NCD plasma.

\subsection{Photon fluence}
We used the yield of \({}^{62}_{29}\text{Cu}\) to determine the $\gamma$ fluence. The \({}^{62}_{29}\text{Cu}\) yield was derived through analyzing the positrons recorded by IP. The number of positrons $N$ can be obtained through Eq. (1) and Eq. (2) \cite{williams2014}, where $G$ is the gray-scale value, $PSL$ is the  photo-stimulated luminescence value, $R$ represents the pixel size of IP, $V$ is the voltage setting used during scanning, $h(V)$ denotes the scanner’s sensitivity function at voltage $V$, and $L$ indicates the range of exposure values, which was set to 5 in our experiment. Additionally, \( g(E) \) represents the response efficiency of the IP to positrons with energy \( E \), \( f(t) \) denotes the signal attenuation coefficient of the IP plate, and \( k(\theta) \) stands for the correction factor for different incident angles of positrons. Under our experimental condition, the emitted positrons have an energy of 1320.7 keV, corresponding to a response efficiency \( g(E) \) = 0.008 \cite{doyama1997}. The measurement was conducted for 15 minutes, resulting in a signal attenuation factor \( f(t) \) = 0.8 \cite{doyama1997}. We adopted the value of \( k(\theta) \) = 1.8 by averaging over the response of the IP to positrons at various incidence angles.
\begin{align}
    PSL &= \left( \frac{G}{2^{16} - 1} \right)^2 \left( \frac{R}{100} \right)^2 h(V) 10^{\frac{L}{2}} 
\end{align}
\begin{align}
    N &= \frac{\text{PSL}}{g(E) \cdot f(t) \cdot k(\theta)}
\end{align}

The total number of positrons $N_{\text{tot}}$ was obtained by summing the contributions from all pixels within the activated area of the copper sheet.
The activity of activated copper sheet immediately following laser shot \(A_0\) was determined using Eq. (3), and the yield of \({}_{29}^{62}\text{Cu}\) \( Y_i \) can be obtained through Eq. (3-5), where $A$ represents the activity at time $t$, \(N_t\) is the number of isotopes remaining existed at time $t$, \(\lambda\) is the decay constant, and \(T_{1/2}\) denotes the half-life of \({}_{29}^{62}\text{Cu}\). Thus, \(\rho\) represents the decay ratio over the measurement interval. Specifically, \(t_1\) represents the time when the copper sheet was placed on the IP after each shot, approximately 12 minutes, while \(t_2\) denotes to the end time of the measurement, corresponding to we removed the sheet from IP, approximately 27 minutes.
\begin{equation}
	\begin{aligned}
         A &= \lambda N_t = A_0 e^{-\lambda  t}, \lambda = \frac{\ln(2)}{T_{1/2}} 
	\end{aligned}
\end{equation}
\begin{equation}
	\begin{aligned}
		\rho &= e^{-\lambda  t_1} - e^{-\lambda  t_2} \\
	\end{aligned}
\end{equation}
\begin{equation}
	\begin{aligned}
        Y_i &= N_{\text{tot}} / \rho \\
	\end{aligned}
\end{equation}
Through Eq. (6), the photon flux of $\gamma$ radiation source $N_{\gamma}(E)$ can be determined from the yield of \({}_{29}^{62}\text{Cu}\), where \( N_T \) is total number of target nuclei, \( N_{\gamma}(E) \) denotes the incident photon flux at energy $E$; \( \sigma_i(E) \) indicates cross section value of \({}^{63}_{29}\text{Cu}\) (\(\gamma, n\)) \({}^{62}_{29}\text{Cu}\) at energy $E$.
\begin{equation}
	Y_i = N_T \int_{E_{\text{th}}}^{\infty} N_{\gamma}(E) \sigma_i(E) \, dE
\end{equation}
For simplifying the calculation, we integrated the energy-dependent cross section of the reaction channel \({}_{29}^{63}\text{Cu} (\gamma, n) \, {}_{29}^{62}\text{Cu}\) over energy and normalized it by the FWHM of the reaction channel shown in Fig. 2 to obtain a constant cross section for the calculation.

The calculated photon fluence of $\gamma$ radiation for shot 102 and shot 105 were shown in Tab. 2.
The photon yield can go up to \( 7.7 \times 10^{10} \) photons$/MeV/sr$. The uncertainty was 11.2\%, which included statistical error, 5\% IP efficiency uncertainty, and a 10\% error associated with the scanning process.
The energy uncertainty is estimated to be 2.5 $MeV$, corresponding to the FWHM of the cross section for the reaction \({}^{63}_{29}\text{Cu}\) (\(\gamma, n\)) \({}^{62}_{29}\text{Cu}\).
Normalized to the laser energy, we obtained a fluence of  \(10^{9}\) $photons/J$, corresponding to the result of  \( 7.7 \times 10^{10} \) photons$/MeV/sr$, which represents a tenfold improvement compared to the result obtained by Stoyer in NOVA laser system, where they measured approximately \(10^{8}\) $photons/J$ at 10-times higher laser intensity \cite{stoyer2001}.

Moreover, it was found that the photon yields and activity were enhanced by shifting the laser focal position 200 \(\mu\)m into the target from the surface while maintaining identical laser parameters and using the same target. 
The observed enhancement due to the change of laser focus position is consistent with the findings reported by Hang and Zhang \cite{hang2021,zhang2020}, where an increase in plasma temperature was achieved by varying the distance between the laser focal point and the target surface.
\begin{table}[ht]
	\caption{Photon yields and activity calculated results comparison between shot 102 and shot 105.}
	\centering
	\begin{tabular}{|c|c|c|c|}
		\hline  
		Shot Number & 102  & 105 \\
        \hline
        Focus position & surface & 200 \(\mu\)m inside \\
		\hline
		Photon yields$/MeV/sr$ & \( 4.6 \times 10^{10} \)  & \( 7.7 \times 10^{10} \) \\
		\hline
		Activity ($Bq$) & \( 3.4 \times 10^{4} \)  & \( 5 \times 10^{4} \) \\
		\hline
	\end{tabular}
	\label{tab:photon_yields_activity}
\end{table}
The activity 385 $Bq/J$, representing a significant improvement compared to previous measurements using a solid target, which yielded 120 $Bq/J$ \cite{norreys1999}. 
\begin{figure}[!htbp]
	\centering
	\includegraphics[scale=.2]{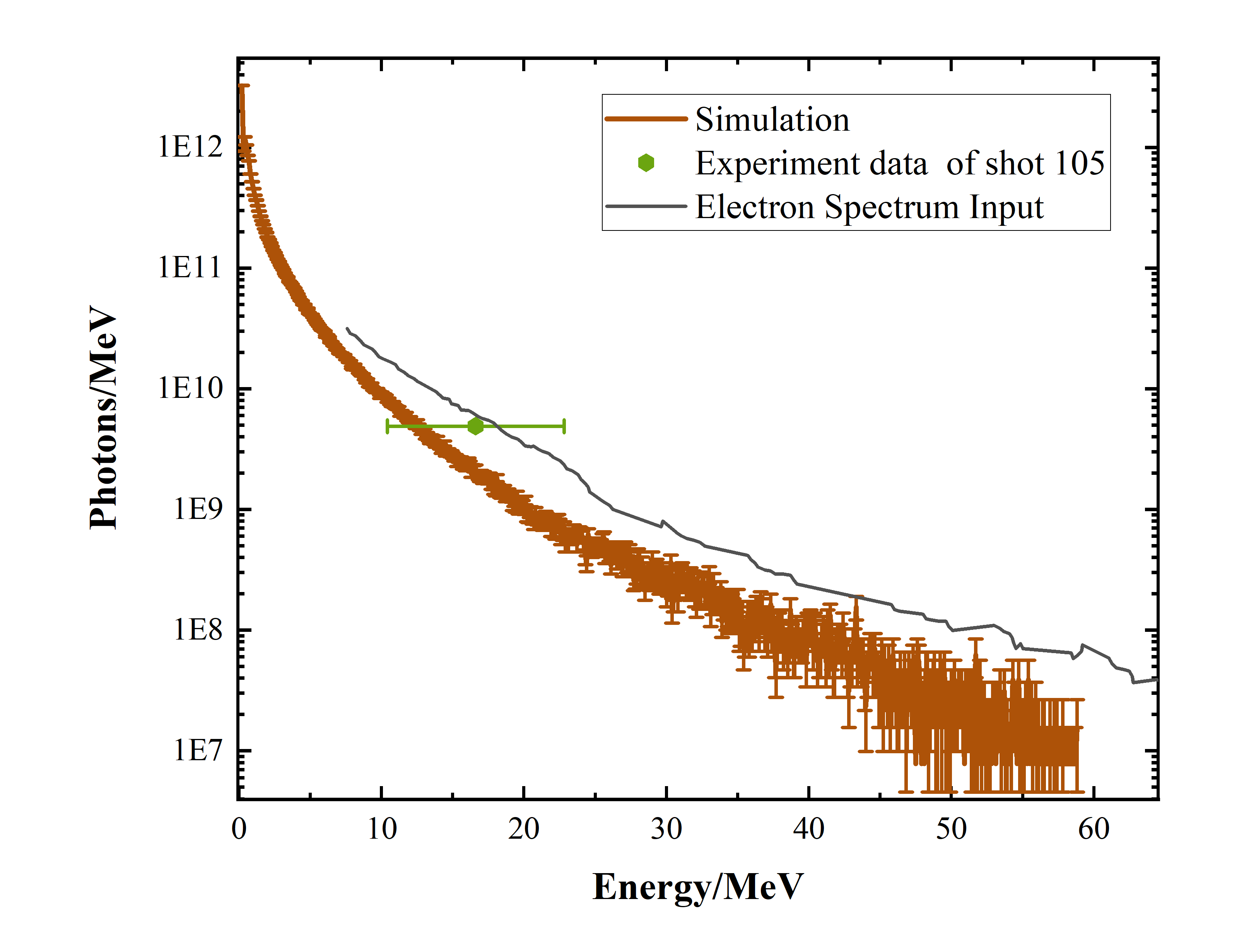}
	\caption{Comparison of the $\gamma$ flux between experimental data and Geant4 simulation result.}
	\label{FIG:5}
\end{figure}

We also performed a GEANT4 (version 11.1.0) simulation\cite{geant4} of bremsstrahlung generation and detected the energy spectrum of the $\gamma$ radiation. In the simulation, we used the QGSP\_BIC\_HP physics list and run a total of \(10^{13}\) electrons to achieve statistical significance. The electron divergence angle and spectrum obtained from the experiment with similar laser parameter and same target was used as the input for the electron source. This electron spectrum follows a Boltzmann distribution with a temperature of 12.5 $MeV$.
These electrons were directed onto a 3 mm-thick tantalum convertor to generate bremsstrahlung radiation, and a virtual detector was positioned behind the convertor to record the resulting $\gamma$ flux. 
As shown in Fig. 5, the simulation results are in good agreement with the experimental data, thereby reinforcing the reliability of our measurements.

\section{Conclusion}
We presented an approach of generating high-brightness and high-energy $\gamma$ radiation source, and diagnosing its divergence angle and fluence in the energy range of 17.5 $\pm$ 2.5 $MeV$.
We employed a nanosecond laser to indirectly heat a low-density foam target, thereby generating a large-scale, slowly evolving NCD plasma. Subsequently, an intense picosecond laser pulse directly interacted with the NCD plasma, producing a high-current, high-energy and collimated electron beam via the DLA mechanism. When these electrons traversed a 3 $mm$ tantalum converter, they generated a well-directed $\gamma$ radiation source through bremsstrahlung.
The $\gamma$ radiation irradiated copper and induced the photonuclear reaction \( _{29}^{63} \text{Cu} \)($\gamma$,n) \( _{29}^{62} \text{Cu} \), producing radioactive isotope \( _{29}^{62} \text{Cu} \). We used IP to record the positrons emitted from the $\beta^+$ decay of \( _{29}^{62} \text{Cu} \). 
We used the spatial distribution and intensity of gray-scale value caused by positrons on IP to characterize the divergence angle and fluence of the  $\gamma$ radiation source.
The half divergence angle of the $\gamma$ radiation source was about 11$^\circ$, and the fluence in the energy range of 17.5 $\pm$ 2.5 $MeV$ was \( 7.7 \times 10^{10} \) $photons/MeV/sr$. 

Good agreement was achieved between experiment results and the Geant4 Monte Carlo simulation here.
This method has proven to be an effective approach for measuring both the divergence angle and fluence of MeV $\gamma$ radiation.
This work paved a way for application in laboratory nuclear physics driven by high power laser at moderate relativistic intensity.

\section{Acknowledgements}
The experiment was conducted at the XG-III facility in Mianyang. The authors would like to express their sincere gratitude to all the staff at the Laser Fusion Research Center. The work was supported by National Key R\&D Program of China, No. 2022YFA1603300, National Natural Science Foundation of China (Grant Nos. 12422512, 12405238, 12120101005 and 12175174), the Postdoctoral Fellowship Program of CPSF under Grant Number GZC20241372, the China Postdoctoral Science Foundation under Grant Number 2024M762569, the Postdoctoral Research Funding Project of Shaanxi Province under 2024BSHYDZZ014.


\bibliographystyle{plainnat}



\end{document}